\documentclass{INTERSPEECH2023}
\interspeechcameraready
\usepackage{multirow, multicol}
\usepackage[nolist]{acronym}
\usepackage{cite}
\usepackage{adjustbox}
\usepackage{listings}
\usepackage{xcolor}

\definecolor{codegreen}{rgb}{0,0.6,0}
\definecolor{codegray}{rgb}{0.5,0.5,0.5}
\definecolor{codepurple}{rgb}{0.58,0,0.82}
\definecolor{backcolour}{rgb}{0.95,0.95,0.92}

\makeatletter
\def\bstctlcite{\@ifnextchar[{\@bstctlcite}{\@bstctlcite[@auxout]}}
\def\@bstctlcite[#1]#2{\@bsphack
  \@for\@citeb:=#2\do{%
    \edef\@citeb{\expandafter\@firstofone\@citeb}%
    \if@filesw\immediate\write\csname #1\endcsname{\string\citation{\@citeb}}\fi}%
  \@esphack}
\makeatother 

\usepackage{inconsolata}
\lstdefinestyle{mystyle}{
    backgroundcolor=\color{backcolour},   
    commentstyle=\color{codegreen},
    keywordstyle=\color{magenta},
    numberstyle=\tiny\color{codegray},
    stringstyle=\color{codepurple},
    basicstyle=\ttfamily\footnotesize,
    breakatwhitespace=false,         
    breaklines=true,                 
    captionpos=b,                    
    keepspaces=true,                 
    numbers=left,                    
    numbersep=5pt,                  
    showspaces=false,                
    showstringspaces=false,
    xleftmargin=14pt,
    framexleftmargin=14pt,
    showtabs=false,                  
    tabsize=2
}

\colorlet{punct}{red!60!black}
\definecolor{background}{HTML}{EEEEEE}
\definecolor{delim}{RGB}{20,105,176}
\colorlet{numb}{magenta!60!black}

\lstdefinelanguage{json}{
    basicstyle=\normalfont\ttfamily,
    numbers=left,
    numberstyle=\tiny\color{codegray},
    xleftmargin=14pt,
    framexleftmargin=14pt,
    stepnumber=1,
    numbersep=8pt,
    showstringspaces=false,
    breaklines=true,
    frame=lines,
    backgroundcolor=\color{backcolour},
    literate=
     *{0}{{{\color{numb}0}}}{1}
      {1}{{{\color{numb}1}}}{1}
      {2}{{{\color{numb}2}}}{1}
      {3}{{{\color{numb}3}}}{1}
      {4}{{{\color{numb}4}}}{1}
      {5}{{{\color{numb}5}}}{1}
      {6}{{{\color{numb}6}}}{1}
      {7}{{{\color{numb}7}}}{1}
      {8}{{{\color{numb}8}}}{1}
      {9}{{{\color{numb}9}}}{1}
      {:}{{{\color{punct}{:}}}}{1}
      {,}{{{\color{punct}{,}}}}{1}
      {\{}{{{\color{delim}{\{}}}}{1}
      {\}}{{{\color{delim}{\}}}}}{1}
      {[}{{{\color{delim}{[}}}}{1}
      {]}{{{\color{delim}{]}}}}{1},
}

\newacro{DER}[DER]{diarization error rate}
\newacro{JER}[JER]{jaccard error rate}
\newacro{EV}[EV]{envelope variance}
\newacro{DNN}[DNN]{deep neural network}
\newacro{WER}[WER]{word error rate}
\newacro{SSL}[SSL]{self-supervised learning}
\newacro{SNR}[SNR]{signal-to-noise ratio}
\newacro{SSLR}[SSLR]{self-supervised learning representation}
\newacro{GSS}[GSS]{guided source separation}
\newacro{SOT}[SOT]{serialized output training}
\newacro{SSE}[SSE]{speech separation and enhancement}
\newacro{SA-WER}[SA-WER]{speaker-attributed word error rate}
\newacro{DA-WER}[DA-WER]{diarization-attributed word error rate}
\newacro{ASR}[ASR]{automatic speech recognition}

\renewenvironment{thebibliography}[1]{
  \begin{oldthebibliography}{#1}
    \setlength{\itemsep}{0em}
    \setlength{\parskip}{0em}
}
{
  \end{oldthebibliography}
}

\title{The CHiME-7 DASR Challenge: Distant Meeting Transcription with Multiple Devices in Diverse Scenarios}
\name{Samuele Cornell$^{1,2}$, Matthew Wiesner$^3$, 
Shinji Watanabe$^2$, Desh Raj$^3$, Xuankai Chang$^2$, Paola Garcia$^3$, Matthew Maciejewski$^3$, Yoshiki Masuyama$^4$, Zhong-Qiu Wang$^2$, Stefano Squartini$^1$, Sanjeev Khudanpur$^3$}
\address{
  $^1$Università Politecnica delle Marche, Italy
  $^2$Carnegie Mellon University, USA \\  
  $^3$Johns Hopkins University, USA 
  $^4$Tokyo Metropolitan University, Japan}
\email{}

\begin{document}
\bstctlcite{IEEEexample:BSTcontrol}
\setlength{\abovedisplayskip}{2pt}
\setlength{\belowdisplayskip}{2pt}



\maketitle
 
\begin{abstract}
The CHiME challenges have played a significant role in the development and evaluation of robust automatic speech recognition (ASR) systems. 
We introduce the CHiME-7 distant ASR (DASR) task, within the 7th CH\-iME challenge. This task comprises joint ASR and diarization in far-field settings with multiple, and possibly heterogeneous, recording devices.  
Different from previous challenges, we evaluate systems on 3 diverse scenarios: CH\-iME-6, DiPCo, and Mixer 6.
The goal is for participants to devise a single system that can generalize across different array geometries and use cases with no a-priori information.
Another departure from earlier CHiME iterations is that participants are allowed to use open-source pre-trained models and datasets. 
In this paper, we describe the challenge design, motivation, and fundamental research questions in detail. 
We also present the baseline system, which is fully array-topology agnostic and features multi-channel diarization, channel selection, guided source separation and a robust ASR model that leverages self-supervised speech representations (SSLR).
\end{abstract}


\section{Introduction}
Despite significant leaps made in the last decade, reliable conversational speech recognition remains a significant challenge~\cite{watanabe2020chime, szymanski2020we}.
This is particularly true in meeting scenarios which are often constrained to use only distant array devices~\cite{watanabe2020chime, chen2020continuous, raj2021integration, Araki2018, Gburrek2022}. 
Besides difficulties due to noise and reverberation in far-field speech, in meeting scenarios a crucial challenge is the presence of overlapped speech, which may amount to more than 15\% of total conversation~\cite{cornell2022overlapped}. 
A third issue are linguistic artifacts arising from informal settings or domain-specific words, an often disregarded characteristic in ASR research, as most commonly used benchmark datasets feature scripted or semi-scripted conversations~\cite{szymanski2020we}. 

As these difficulties arise from diverse factors, multi-talker distant conversational speech recognition requires a comprehensive approach for the development and integration of both front-end and ASR back-end techniques.
Over the years, several challenges and corpora have played a significant role in advancing research in this field. These include, but are not limited to, ASpiRE~\cite{harper2015automatic}, AMI~\cite{carletta2005ami}, ICSI~\cite{janin2003icsi}, the CH\-iME (1--4) challenges~\cite{barker2013pascal, vincent2013second, barker2015third, vincent20164th}, the Sheffield Wargames corpus~\cite{fox2013sheffield}, and DIRHA~\cite{cristoforetti2014dirha}. 
More recently, we have seen VOiCES~\cite{richey2018voices}, DiPCo~\cite{van2019dipco}, LibriCSS~\cite{chen2020continuous}, Alimeeting~\cite{yu2022m2met}, Ego4D~\cite{grauman2022ego4d}, and the recent CH\-iME 5 and 6 challenges~\cite{barker2018fifth, watanabe2020chime}. 

Several of these datasets/challenges, such as DIRHA, VOiCES, and CHiME-4, focused primarily on acoustic robustness, thus featuring pre-segmented single speaker utterances in a noisy/reverberant environment captured by multiple microphones. The objective in this case was to foster research in robust ASR methods, ignoring possible issues that can arise from overlapped speech, wrong segmentation, and colloquial language. These datasets comprised either fully simulated or semi-real environments, where utterances come from an audiobook speech corpus such as LibriSpeech~\cite{panayotov2015librispeech}. 
A notable exception in this regard is LibriCSS. It features a simulated, unsegmented full meeting scenario between multiple participants in semi-real conditions; it was obtained by playing LibriSpeech utterances via loudspeakers in a real room. 

AMI, ICSI, and Sheffield Wargames were historically among the first projects that featured real unsegmented meeting-style scenarios recorded in far-field settings. 
Although these are more expensive to collect compared with simulated or semi-real datasets, they better reflect possible real-world applications. 
Recently, this line of research has seen renewed interest, and has resulted in the collection and creation of datasets such as CH\-iME-5/6, DiPCo, Alimeeting, and Ego4D. The progress made in ASR, diarization and speech separation is a significant contributing factor to this trend, and it suggests that reliable and robust transcription in these scenarios is within our reach in the coming years.  

The proposed CH\-iME-7 DASR challenge\footnote{CHiME-7 DASR website: \href{https://www.chimechallenge.org/current/task1/index}{chimechallenge.org/current/task1/index}} follows this latter line of research on real, unsegmented meeting scenarios. Our main goal is to foster research in an important direction: how can we build systems that can generalize across a \textit{wide range of real-world settings} and provide reliable ASR performance under adverse acoustic conditions? 
Additionally, we aim to incorporate recent progresses in \textit{self-supervised learning} to address this problem as well as supervised DNN-based speech separation and enhancement (SSE), by allowing the use of external data and pre-trained models. 

\section{Motivation}
\label{sec:motivation}

Compared to the previous CH\-iME editions, CH\-iME-7 DASR features three main novelties, motivated by recent promising directions in speech processing.

\subsection{Diverse Scenarios}
\label{sec:motivation_diverse}

To expand the breadth of evaluation conditions, we include two additional datasets (along with CHiME-6) --- DiPCo~\cite{van2019dipco}, and Mixer 6 Speech~\cite{brandschain2010mixer}. 
The objective is to encourage research towards methods that (1) work well across different array topologies: linear (CH\-iME-6), circular (DiPCo), or heterogeneous (Mixer 6); (2) are capable of handling variable numbers of speakers in each session; (3) account for linguistic differences between dinner party scenarios (CH\-iME-6 and DiPCo) versus interviews (Mixer 6); and (4) can effectively handle diverse acoustic conditions.
Such variability reflects real use-cases where such cross-domain generalization capability is highly desirable. 



\subsection{``Foundation'' Models} 
Pre-trained speech models trained on large datasets are increasingly becoming available to the public. These include supervised models such as OpenAI Whisper~\cite{whisper}, Nvidia JASPER~\cite{Li2019JasperAE} and QuartzNet~\cite{Kriman2019QuartznetDA}, and self-supervised ones such as Wav2Vec 2.0~\cite{baevski2020wav2vec}, HuBERT~\cite{hsu2021hubert}, and WavLM~\cite{chen2022wavlm}. 
The latter, in particular, have proved effective in many downstream applications \cite{yang2021superb}, and their integration with front-end speech enhancement has enabled state-of-the-art performance on benchmarks such as CHiME-4~\cite{chang2022end, masuyama2022end}. 
Leveraging these models offers unique opportunities as well as novel challenges. By allowing the use of pre-trained models, the CHiME-7 DASR task becomes more accessible to participants with limited resources, since training is often more efficient and quicker with such models. While SSLR models can be quite large, the features could be pre-extracted, thus enabling significant computational savings in the training phase.
From a research perspective, it may be interesting to investigate how such models, which are trained in monoaural conditions, can make effective use of multi-channel audio and/or can be distilled into smaller models.


\subsection{Open-Source Datasets}
The allowable external data sources are expanded to include many commonly used open-source datasets (e.g. LibriSpeech~\cite{panayotov2015librispeech} and FSD50k~\cite{fonseca2021fsd50k}). They facilitate the use of models for which, in prior challenges, necessary training data were lacking, but also enables the study of new research directions.
For example, participants can create large synthetic datasets to train powerful \ac{DNN}-based \ac{SSE}~\cite{kanda2022vararray, yoshioka2022vararray}. The structure of the challenge then encourages research into whether such models can effectively help improving real-world meeting transcription.  
This latter research direction also intersects with this year's ``sister'' CH\-iME-7 UDASE challenge, whose focus is on unsupervised domain adaptation for speech enhancement. 




\section{Challenge Tracks \& Rules}
\label{sec:rules}

The CH\-iME-7 DASR challenge features two tracks --- a main track, and an optional sub-track, as described below.


\subsection{Main Track}
\label{sec:main_track}

\begin{figure}
    \centering
    \includegraphics[width=\linewidth]{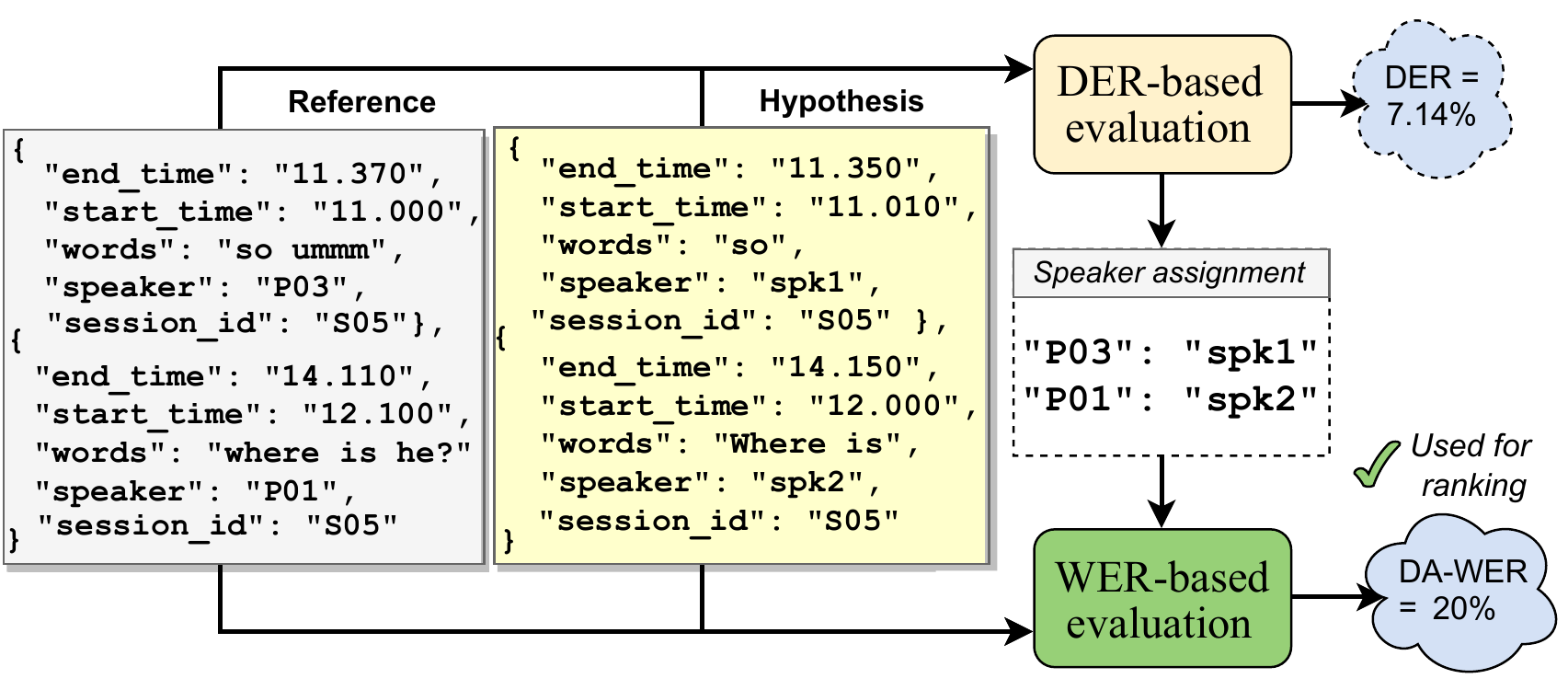}
    \caption{Evaluation scheme for CHiME-7 DASR main track. Optimal speaker assignments are first determined using a DER-based evaluation, and then used to compute SA-WER, which is used as the final ranking metric.}
    \label{fig:main_eval}
    \vspace{-1.7em}
\end{figure}

For the main track, we provide multi-channel recordings for the whole session without segmentation or speaker labels, and participants are required to generate time-marked, speaker-attributed transcripts, as shown in Fig.~\ref{fig:main_eval}. For any session, let $\mathbf{r}_i = (\Delta, s, \mathbf{v})$ denote a time-marked, speaker-attributed transcript of a segment, where $\Delta = (t_{\mathrm{st}}, t_{\mathrm{en}})$ is a tuple containing start and end times, $s \in \mathbb{Z}^+$ is the speaker label\footnote{Actual label names may be arbitrary, but can be mapped to the set of positive integers without loss of generality.}, and $\mathbf{v} \in \Sigma^*$ represents the transcript for the segment, for some vocabulary $\Sigma$. A session is then completely defined by $\mathbf{R} = \{\mathbf{r}_1,\ldots, \mathbf{r}_N\}$, which we call the \textit{reference}. Challenge participants are required to estimate the reference through generated \textit{hypothesis}, denoted by $\mathbf{H} = \{\mathbf{h}_1,\ldots,\mathbf{h}_{\hat{N}}\}$. 

In this challenge, estimated speaker labels $\hat{s}$ are not required to be identical to the reference labels. This flexibility poses a challenge in computing speaker-attributed word error rates (SA-WER) since, unlike earlier studies~\cite{Kanda2022StreamingMA}, the mapping between reference and estimated labels is not known.

We solve this problem using mechanisms employed in the evaluation of diarization systems, as shown in Fig.~\ref{fig:main_eval}. Let $\Psi: \mathbb{Z}^+\rightarrow \mathbb{Z}^+\cup\{\phi\}$ denote a mapping from the reference labels to the estimated labels. We use the Hungarian method~\cite{Toroslu2007IncrementalAP} to obtain the optimal mapping $\hat{\Psi}$ that minimizes the diarization error rate (DER) between the reference segments and hypothesized segments, i.e.,
\begin{equation}
    \hat{\Psi} = \arg \min_{\Psi} \text{DER}(\mathbf{R}, \mathbf{H}),
\end{equation}
where the DER is calculated with a $250$\,ms collar, following the setup in CH\-iME-6. Given $\hat{\Psi}$, we compute this newly proposed diarization attributed WER (DA-WER) metric as
\begin{equation}
    \text{DA-WER}(\mathbf{R},\mathbf{H}) = \frac{\sum_{s}\mathcal{L}(\frown \mathbf{R}^s,\frown \mathbf{H}^{\hat{\Psi}(s)})}{\sum_s| \frown \mathbf{R}^s |},
\end{equation}
where $\mathbf{R}^s$ denotes segments in $\mathbf{R}$ with speaker $s$, $\mathcal{L}$ is the Levenshtein distance, and the $\frown$ operator concatenates segment transcripts of its operand. Final ranking of systems will be based on the DA-WER metric macro-averaged across all scenarios. 
The macro-average operation is to encourage participants to develop a model whose performance is consistent across all three scenarios. 
Note that this evaluation is different from the concatenated minimum-permutation WER (cpWER) based ranking in the CHiME-6 challenge, since it requires participants to also produce estimates of the utterances boundaries. 
Utterance boundaries estimation are important in actual applications, e.g. for double checking the transcripts. 
It also differs from the Asclite multi-dimensional Levenshtein metric~\cite{fiscus2006multiple}, which can account for segmentation errors when the time-based cost is used. Contrary to this latter, DA-WER does not require word-level alignment, which is quite time-consuming to double-check for ground-truth purposes. DA-WER instead offers a simple and ``loose'' way to encourage participants to produce reasonable segmentation at the utterance level, which is sufficient in most applications. 
Future work is needed to devise more principled metrics for joint ASR and diarization. 

\subsection{Optional Sub-Track}

Similar to the CH\-iME-6 challenge, we include an optional sub-track where participants can use oracle diarization. The goal of this sub-track is to assess the acoustic front-end and ASR performance and disentangle it from the impact of diarization. The segmentation and speaker labels are known and are freely usable by participants, here the problem reduces to a standard ASR task.


\subsection{Rules}
Detailed description of the task rules are available in the CH\-iME-7 DASR website\footnote{\href{https://www.chimechallenge.org/current/task1/rules}{chimechallenge.org/current/task1/rules}}. Hereafter we briefly summarize them.
\begin{itemize}[noitemsep,topsep=0pt,leftmargin=*]
\item Only some external data and pre-trained models are allowed. Participants are encouraged to propose additional ones in the first month.
\item Close-talk microphones cannot be used for inference.
\item \textit{Automatic domain identification is prohibited}. This includes the use of a-priori information such as array topology (including number of channels). \textit{A single system must be submitted for all three scenarios.}
\item Automatic channel selection is allowed and encouraged.
\item \textit{Self-supervised adaptation/training is allowed}, as long as it is performed for each evaluation session independently i.e. as it would happen in a real-world deployment.
\item Participants may use all available data (including external data) for language model training, but cannot use large language models (LM) such as BERT~\cite{kenton2019bert}. This because we want participants to focus mainly on acoustic robustness.
\end{itemize}


\section{Datasets}
\label{sec:dataset}

As explained in Section~\ref{sec:motivation}, CH\-iME-7 DASR is composed of three different scenarios/datasets: CH\-iME-6, DiPCo, and Mixer 6 Speech. 
For the purpose of this challenge, several changes have been made to these aforementioned data sources. In particular, we partially re-annotated Mixer 6 and we harmonized the transcripts across each of the three by using the same convention for common non-verbal speech sounds such as ``mhm'', ``mmh'', ``hmm'', which are rather frequent. Dataset statistics are summarized in Table~\ref{tab:class_stats}.
The annotation comes in the form of JSON files (one for each session), in the style of CH\-iME-6 (see Listing~\ref{lst:annotation}). %
Hereafter, we describe each scenario in more detail. 

\begin{table}[h]
\centering
\footnotesize
\caption{CH\-iME-7 DASR data overview for the three scenarios. For the CHiME-6 data, we show statistics for the original annotation, as well as a force-aligned (FA) annotation. 
We report the number of utterances, speakers, and sessions, as well as silence (sil), single-speaker speech (1-spk) and overlapped speech (ovl) ratios over the total duration. 
}\label{tab:class_stats}
\adjustbox{max width=\linewidth}{
\begin{tabular}{@{}llrrrrrrr@{}}
\toprule   
 \textbf{Scenario} & \textbf{Split} & \textbf{Size (h)} & \textbf{Utts} & \textbf{Spk.} & \textbf{Sess.} & \textbf{sil (\%)} & \textbf{1-spk (\%)} & \textbf{ovl (\%)}   \\
\midrule
\multirow{3}{*}{\textbf{CH\-iME-6}} & \multirow{1}{*}{train} &  \multirow{1}{*}{30:57} &  67615  & \multirow{1}{*}{28} & \multirow{1}{*}{14} & 22.6  & 53.5  & 23.9  \\
& \multirow{1}{*}{dev} &  \multirow{1}{*}{4:27} &  7437  & \multirow{1}{*}{8} & \multirow{1}{*}{2} & 12.5  & 43.7  & 43.8  \\
& \multirow{1}{*}{eval} &  \multirow{1}{*}{10:21} &  20683  & \multirow{1}{*}{12} & \multirow{1}{*}{4} & 19.9  &  50.9  &  29.2  \\
\midrule
\multirow{3}{*}{\begin{tabular}{@{}l@{}}\textbf{CHiME-6}\\\,\,\,\,\,\,\,\textbf{(FA)}\end{tabular}} & \multirow{1}{*}{train} &  \multirow{1}{*}{30:57} &  78640  & \multirow{1}{*}{28} & \multirow{1}{*}{14} & 38.8  & 50.2  & 11.0  \\
& \multirow{1}{*}{dev} &  \multirow{1}{*}{4:27} &  11020  & \multirow{1}{*}{8} & \multirow{1}{*}{2} & 24.1  & 54.2  & 21.7  \\
& \multirow{1}{*}{eval} &  \multirow{1}{*}{10:21} &  81890  & \multirow{1}{*}{12} & \multirow{1}{*}{4} & 38.8  &  49.2  &  12.0  \\
\midrule
\multirow{2}{*}{\textbf{DiPCo}} & dev &  2:43  &  3673  & 16 & 5 & 7.6 &  66.5 &   25.8   \\
& eval &  2:36 & 3405  & 16 & 5 &  9.2 & 65.8 & 25.0 \\
\midrule
\multirow{4}{*}{\textbf{Mixer 6}}  & train calls &  36:09 & 27280  & 81 & 243 & --   & --  &    --  \\
 & train intv & 26:57  &  29893  & 77 & 189 & -- &  -- &  --    \\
& dev &  14:75 &   14863 & 27 & 59 & 3.1 & 76.2  & 20.7     \\
& eval & 5:45 & 5115  & 18 & 23 &  2.4 & 83.6 & 13.9 \\
\bottomrule
\end{tabular}
}
\vspace{-2em}
\end{table}

\subsection{CH\-iME-6}
\label{sec:chime6}
The CHi\-ME-6  dataset features real dinner parties with 4 participants in a home environment, usually across different rooms. Due to the particular setting, it features informal speech and low \ac{SNR}. Recordings from binaural microphones worn by each speaker are provided along with distant speech captured by 6 Kinect array devices with 4 microphones each for a total of 24 microphones. Two annotations are provided~\cite{watanabe2020chime}, one manual and another obtained via forced alignment (FA)\footnote{\href{https://github.com/chimechallenge/CHiME7\_DASR\_falign}{chimechallenge/CHiME7\_DASR\_falign}}.  
In CHi\-ME-7 DASR, we retain the CH\-iME-6 dataset in its original form, with the exception that two sessions were moved from the original \texttt{train} set to the \texttt{eval} set. This in order to have a bigger \texttt{eval} set, with more variety in acoustical conditions, compared to the CHi\-ME-6 challenge. Importantly, we can still reasonably compare results with the previous challenge ones, on the \texttt{dev} set and the original portion from \texttt{eval}. 
We also provide universal evaluation map (UEM) files to exclude the speaker enrollment from evaluation, a feature missing from the previous iteration which incorrectly scored this portion as silence.

\begin{lstlisting}[language=json, firstnumber=1, basicstyle=\footnotesize, caption=Sample annotation for the CHiME-6 dataset., label={lst:annotation}]
``end_time'': ``76.340'',
``start_time'': ``73.500'',
``words'': ``Let's do lunch. [laughs]'',
``speaker'': ``P08'',
``ref'': ``U02'',
``location'': ``kitchen'',
``session_id'': ``S02''
\end{lstlisting}


\subsection{DiPCo}
\label{sec:dipco}

DiPCo consists of 10 recorded dinner party sessions, each between 4 speakers, recorded on 5 far-field devices each with a 7-mic circular array (six plus one microphone at the center).
Per-speaker close-talk microphones are provided, and have less cross-talk than the corresponding mics from CH\-iME-6 . 
Additionally, each session has a lower duration (15--47 mins) and is recorded in the same single room. Importantly, the additional DiPCo data allows us to evaluate participants' submissions in a ``best-case scenario'': where the acoustic conditions between training and testing are reasonably matched.

\subsection{Mixer 6 Speech}
\label{sec:mixer6}

The Mixer 6 Speech set (LDC catalog no. LDC2013S03) features a very different but relevant multi-speaker, multi-channel scenario.
It consists of 594 distinct native English speakers participating in a total of 1425 sessions, each lasting approximately 45 minutes. Sessions include an interview between an ``interviewer'' and a ``subject'' (15 min.), a telephone call (10 min.), and prompt reading. 
Each session is recorded using 14 microphones of varying styles, placed in various locations around one of 2 different rooms (\texttt{LDC} and \texttt{HRM}). 

We make use of the interview and call portions of 450 of the 1425 sessions for which we have collected human-annotated transcriptions. The remaining audio in the 450 sessions, for which no human annotation is available, can be used by the participants, for instance, to support unsupervised or self-supervised training. The 450 sessions are split into \texttt{train}, \texttt{dev}, and \texttt{eval} partitions with no speaker overlap. We only use 13 of the channels in the challenge as the SNR of the 14th channel is too low to be useful (or intelligible).

Human-annotated time-marks and transcripts are available only for the ``subject'', who is usually the dominant speaker. For the ``interviewer`` segments, preliminary transcripts were generated by running Whisper~\cite{radford2022robust} large inference on the lapel microphone (\texttt{eval}), or via manual annotation (\texttt{dev}). Interviewer transcripts were then force-aligned with the audio from the interviewer's lapel microphone (for \texttt{dev}) and/or manually corrected (for \texttt{eval}) to get the ground-truth annotations. 
Information about the splits is shown in Table \ref{tab:class_stats}.
Note that \texttt{eval} was recorded in the (\texttt{HRM}) room; this new room condition tests robustness to shifts in the recording room, while performance on \texttt{dev} represents a more matched scenario. Close-talk channels are not provided to challenge participants for \texttt{eval}.

\section{Baseline Description}
\label{sec:baseline}

\begin{figure}[h]
  \centering  
  \includegraphics[width=0.45\textwidth]{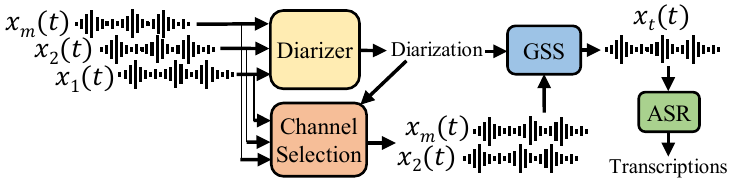}
  \caption{Block scheme for the CH\-iME-7 DASR baseline. The diarizer module is omitted in the acoustic robustness sub-track.}
  \label{fig:chime7_dasr_baseline}
  \vspace{-1.0em}
\end{figure}

The CH\-iME-7 DASR baseline model is shown in Fig.~\ref{fig:chime7_dasr_baseline}. 
For the main track, it features an \textit{array topology agnostic} speech processing pipeline composed of multi-channel diarization, channel selection, \ac{GSS}~\cite{boeddeker2018front}, and monaural ASR. 

For the optional sub-track, the diarizer component is omitted since participants are provided oracle segmentation. 
This baseline system, along with the data preparation, is implemented within ESPNet2~\cite{watanabe2018espnet} as a recipe\footnote{\href{https://github.com/espnet/espnet/tree/chime7task1\_diar/egs2/chime7\_task1/diar\_asr1}{espnet/tree/chime7task1\_diar/egs2/chime7\_task1/diar\_asr1}}. We now describe each component in detail. 

\subsection{Front-End Processing}
\label{sec:front-end}
The baseline system for the CHiME-6 challenge comprised \ac{GSS} for front-end processing~\cite{boeddeker2018front}, with array topology aware choices such as the use of outer microphones. Since CHiME-7 DASR forbids the a-priori knowledge of array geometry, we instead use automatic channel selection using an \ac{EV} approach as proposed in~\cite{WOLF_EV_2014}. 
This measure, while simple, has been found to be effective on CH\-iME-6 data~\cite{cornell2021learning} where it compared favourably with more elaborated data-driven approaches. Channel selection is used to select the most promising subset of microphones for each utterance for later processing via \ac{GSS}.
We use a GPU-accelerated implementation of \ac{GSS}~\cite{Raj2022GPUacceleratedGS} that allows to significantly speed up the inference time (e.g. up to 300x on CHiME-6 development data with two NVIDIA A100 40GB GPUs).

\subsection{Diarization}
During preliminary experiments with the Pyannote~\cite{bredin2020pyannote} diarization pipeline\footnote{\href{https://huggingface.co/pyannote/speaker-diarization/tree/main}{pyannote/speaker-diarization}} on CH\-iME-6, we found that \ac{DER} between microphones belonging to the same array may vary by up to 10\%. This agrees with similar observations made during the CHiME-6 challenge, where teams employed channel fusion methods to subvert these effects~\cite{Arora2020TheJM}. 
At the same time, low speaker variability in CH\-iME-6 and the lack of train segmentation for the interviewer segments in Mixer-6 prohibit methods such as EEND-VC~\cite{kinoshita2021integrating} and suggest the use of embedding-based approaches, unless one wants to create extensive multi-channel synthetic data. 

\begin{figure}[h]
  \centering  
  \includegraphics[width=0.45\textwidth]{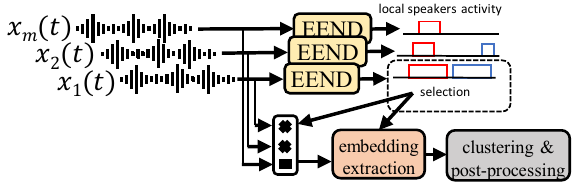}
  \caption{Baseline multi-channel diarization pipeline, built over the Pyannote speaker diarization pipeline.}
  \label{fig:diarization}
  \vspace{-1.5em}
\end{figure}

For these reasons, here we opt for a simple but reasonably effective approach built over the Pyannote~\cite{bredin2020pyannote} diarization system, which is very popular and has been proven to be successful on a wide variety of scenarios.
Our approach is depicted in Figure~\ref{fig:diarization}.
In detail, instead of channel ensembling, we instead leverage the local end-to-end diarization (EEND)~\cite{fujita2019end, bredin2021end} module in the pipeline~\cite{bredin2020pyannote} for channel selection. The maximum number of local speakers is fixed to 3 and we use 5\,s segments.
Only this latter is ran over all channels and, assuming it is robust against false alarms, we select the channel which has the most speech activity as the best channel. 
The rest of the pipeline (including embedding extraction and clustering, which make the bulk of computation) is then run only over this selected microphone channel, leading to a significant speed up versus running the whole pipeline over all channels.
This approach has obviously its limitations, e.g. for two concurrent speakers it may be preferable to select a different microphone for each. Despite this limitation, we found it to work very well on the \texttt{dev} set, significantly outperforming all channels ensembling via DOVERLap~\cite{raj2021dover} (40\% versus 52\% DER on CH\-iME-6 \texttt{dev} set). 
This said, this simple approach can be easily extended in the future to use the top-K channels instead of just one in the clustering phase. 

In the baseline recipe, we fine-tune the local EEND model of the Pyannote pipeline using all CH\-iME-6 \texttt{train} set (by using all far-field channels only) and use for validation Mixer 6 \texttt{dev} set; only this latter is used because Mixer 6 is the most acoustically different dataset among the three scenarios from CHiME-6, so it made sense to use it as validation, to enforce early-stopping and preventing overfitting. 
To make the output of the diarization more suitable for ASR we post-process the diarization output by merging segments from the same speaker which are 0.5 seconds apart. 
The pre-trained model is made available via HuggingFace\footnote{\href{https://huggingface.co/popcornell/pyannote-segmentation-chime6-mixer6}{popcornell/pyannote-segmentation-chime6-mixer6}}.

\subsection{Automatic Speech Recognition}
For the baseline ASR, we directly take the model from \cite{chang2022end, masuyama2022end}, which consists of a hybrid CTC/Attention transformer~\cite{vaswani2017attention} encoder-decoder ASR model with WavLM-based features and has been proven very effective on noisy-reverberant data. 
It is trained on the full CH\-iME-6 \texttt{train} (including the binaural microphones) and the Mixer 6 \texttt{train}. In addition, we used the augmentation scFheme as described in the CHiME-6 baseline~\cite{watanabe2020chime}, which leveraged close-talk microphones and external datasets for RIRs and noises, namely SLR26~\cite{ko2017study} and MUSAN~\cite{snyder2015musan}. We also used \ac{GSS}-enhanced data obtained from the whole CH\-iME-6 training set. Both these addition were found critical to reach state-of-the-art performance.
The training takes approximately 18 hours using two A100 GPUs on a DGXA100 machine.
This model is made available via HuggingFace\footnote{\href{https://huggingface.co/popcornell/chime7\_task1_asr1\_baseline}{popcornell/chime7\_task1\_asr1\_baseline}}
%

\section{Results \& Discussion}
\label{sec:results}

\subsection{Front-End Processing}
\begin{figure}[t]
  \centering  
  \includegraphics[width=0.47\textwidth]{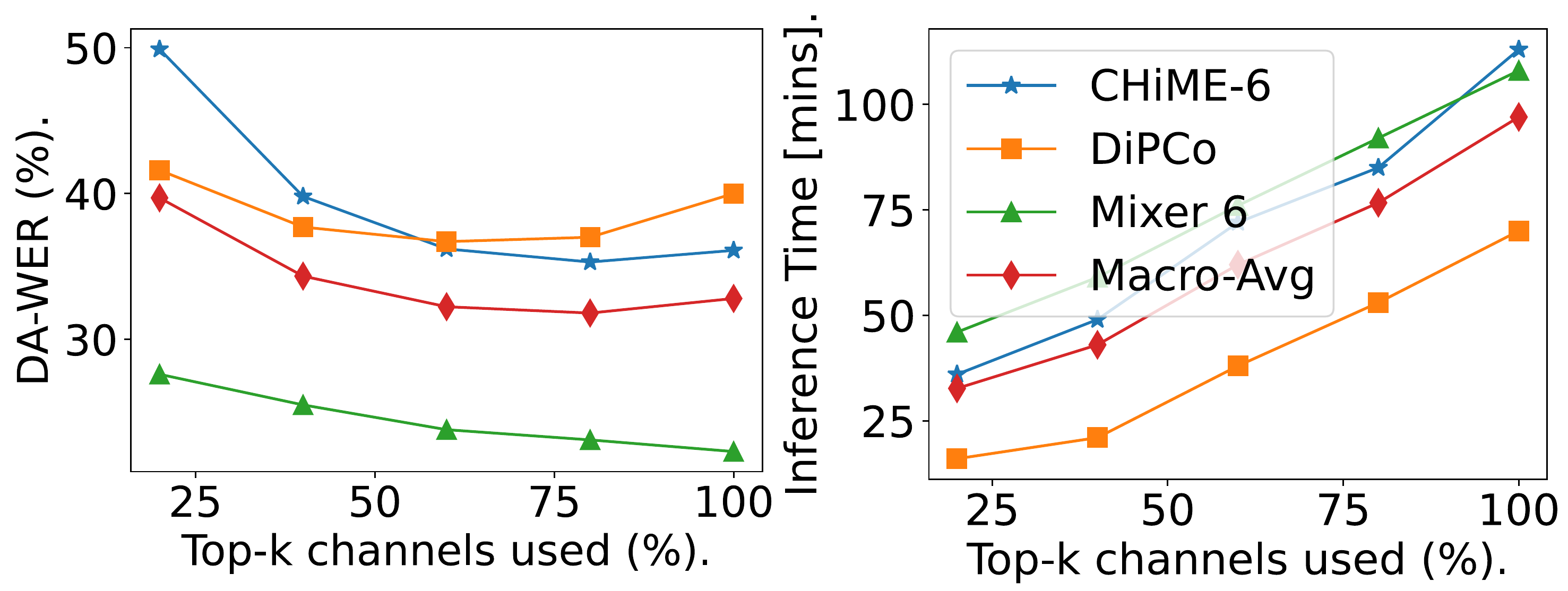}
  \caption{Effect of \ac{EV}-based channel selection with \ac{GSS} on WER performance (left) and inference time (right), on the \texttt{dev} set. We show scenario-wise results and the macro-average.} 
  \label{fig:chime7dasr_selection_dev}
  \vspace{-1.5em}
\end{figure}

As mentioned in Sec.~\ref{sec:front-end}, a key difference between the front-end module in CHiME-7 DASR, compared with the CHiME-6 challenge, is the use of automatic channel selection.
In Fig.~\ref{fig:chime7dasr_selection_dev},
we show WER results and inference time on the \texttt{dev} sets for different selection ratios when the ASR is trained using data containing GSS performed with all channels. 

We can observe two trends. 
First, the time employed to perform selection+\ac{GSS} appears to be approximately linear with the number of microphones used. 
Secondly, the error rate generally decreases as the number of microphones used increases.
Thus there appears to be a trade-off between computational complexity and performance. 
In particular, the best performance for CH\-iME-6 and DiPCo is achieved by using the top 80\%, while Mixer 6 will benefit from the use of all microphones. This could be related to the fact that Mixer 6 is less noisy and all channels might contribute significantly to improving the GSS result. 
In the baseline we always use the top 80\% channels for GSS for both training and inference, as this seems to maximize the performance overall. 

\subsection{Diarization}

In Table~\ref{tab:diarization} we report the results achieved by our proposed diarization system in terms of DER and Jaccard error rate (JER) with $250$\,ms collar. 
We can see that, overall, the system, despite being fine-tuned only on CH\-iME-6 data, generalizes well to the other scenarios. 

\begin{table}[h]
    \caption{CHiME-7 DASR diarization baseline results.  We show scenario-wise results as well as the macro-average.}
    \vspace{-0.2cm}
    \footnotesize
    \label{tab:diarization}
    \centering
    \begin{tabular}{@{}lcccc@{}}
    \toprule
         \multirow{2}{*}{\textbf{Scenario}} & \multicolumn{2}{c}{\textbf{Dev}} &   \multicolumn{2}{c}{\textbf{Eval}} \\
         \cmidrule(r{2pt}){2-3} \cmidrule(l{2pt}){4-5}
         & DER & JER  & DER  & JER \\       
        \toprule
         CH\-iME-6 & 40.0  & 51.1 & 56.3 & 62.5 \\
        DiPCo & 29.8 & 41.4 & 27.9 &  40.9 \\
        Mixer 6 & 16.6 & 22.8  & 9.3 & 11.0  \\ 
        \midrule
        Macro & 28.8 & 38.5 &  31.2 & 38.2 \\  
    \bottomrule
    \end{tabular}
\end{table}

CH\-iME-6 remains by far the most difficult scenarios among the three for what regards diarization, due to the challenging acoustical conditions and frequent overlapped speech. 

\subsection{Overall Results}
In Table~\ref{tab:final_results} we present DA-WER figures obtained on the three CH\-iME-7 DASR scenarios, for both sub-track (\texttt{sub}) and main-track (\texttt{main}), together with their macro averages. 
We compare our baseline model against Whisper~\cite{whisper} large --- note that Whisper cannot be used by participants as it was not included among the allowed pre-trained models (its training data may contain our evaluation).
Nevertheless, it serves as an interesting and strong point of comparison. 

\begin{table}[!htbp]
    \caption{CHiME-7 DASR sub-track results obtained using the best configuration from development set (80\% channels GSS). Whisper results are reported for reference.}
    \vspace{-0.2cm}
    \label{tab:final_results}
    \centering
    \footnotesize
    \setlength{\tabcolsep}{1.5pt}
    \begin{tabular}{clcccc}
    \hline
         \textbf{ASR model} & \textbf{Scenario} & \multicolumn{2}{c}{\textbf{Dev}} &   \multicolumn{2}{c}{\textbf{Eval}} \\
         & & \multicolumn{2}{c}{DA-WER (\%)}  &   \multicolumn{2}{c}{DA-WER(\%)}   \\     
         &  & sub & main  &  sub & main \\ 
        \toprule
         \multirow{4}{*}{Baseline} & CH\-iME-6 & 32.6  & 62.4 & 35.5 & 77.4  \\
                                &  DiPCo & \textbf{33.5}   & 56.6 &  36.3   & 54.7 \\
                                & Mixer 6 & \textbf{20.2}   & \textbf{22.5} & 28.6  & \textbf{33.7} \\ 
                                 \cmidrule{2-6}
                                & Macro & \textbf{28.8} & 47.2 & 33.4  & 55.3  \\ 
    \cmidrule{1-6}
    \multirow{4}{*}{Whisper} & CH\-iME-6 &  \textbf{30.9}   & \textbf{58.4} & \textbf{36.6}  & \textbf{74.0} \\
                                &  DiPCo & 34.5  & \textbf{52.5} & \textbf{35.7}  & \textbf{49.7} \\
                                & Mixer 6 &  21.2   & 23.7 & \textbf{25.2}  & 35.8 \\ 
                                 \cmidrule{2-6}
                                & Macro & \textbf{28.8}  & \textbf{44.8} & \textbf{32.5} & \textbf{53.2}  \\                                 
    \bottomrule
    \end{tabular}
\end{table}

Overall, our baseline ASR model compares favorably with Whisper large especially in the acoustic robustness sub-track where the final macro-average DA-WER scores are very close.
In the main track the difference is more pronounced, and Whisper appears slightly more robust to segmentation errors, due to its significantly larger training data. 

By comparing sub and main columns in Table \ref{tab:final_results}, we can observe that diarization has a big impact on the final recognition score. We discovered that the permutation was always assigned correctly in these results during DA-WER computation, yet, sub-optimal segmentation and mis-attributed sentences degraded considerably the performance, especially on CHiME-6. On the other hand, we can see how both Whisper and the baseline system fail, even with oracle diarization, to reach satisfactory performance suitable for real-world applications, even on the arguably easier Mixer 6 scenario. 



\vspace{-0.2cm}
\section{Conclusions}
In this paper, we introduced the CH\-iME-7 DASR challenge, which builds upon the past CH\-iME-6 and extends it to multiple scenarios which feature diverse array topologies, number of participants, acoustical conditions and linguistic differences, to test meeting transcription systems cross-domain generalization capability. 
Other novelties regard the use of a novel DA-WER metric which also accounts implicitly for diarization accuracy and the allowance for pre-trained ``foundation'' models and popular external datasets. Despite the efforts in developing a solid baseline system, the proposed challenge scenario still poses significant obstacles in achieving accurate transcriptions; this is also demonstrated by results with Whisper large and oracle diarization. Our hope is that participants will devise novel and more effective ways to address this important problem. 

\section{Acknowledgements}
We would like to thank Marc Delcroix, Naoyuki Kamo, Christoph Boedekker and Thilo von Neumann for their help and feedback. 
S. Cornell was partially supported by Marche Region within the funded project ``Miracle'' POR MARCHE FESR 2014-2020.
D. Raj was partially funded by NSF CCRI Grant No. 2120435, and a fellowship from Amazon via the JHU-Amazon Initiative for Interactive Artificial Intelligence (AI2AI).

\small
\bibliographystyle{IEEEtran}
\bibliography{mybib}

\end{document}